# Melatect: A Machine Learning Approach For Identifying Malignant Melanoma In Skin Growths

Vidushi Meel
MIT Kavli Institute, Lexington High School
Boston, MA
meel@space.mit.edu

Asritha Bodepudi
MIT Kavli Institute, Lexington High School
Boston, MA
bodepudi@space.mit.edu

*Abstract*— Malignant melanoma is a common skin cancer that is mostly curable before metastasis -when growths spawn in organs away from the original site. Melanoma is the most dangerous type of skin cancer if left untreated due to the high risk of metastasis. This paper presents Melatect, a machine learning (ML) model embedded in an iOS app that identifies potential malignant melanoma. Melatect accurately classifies lesions as malignant or benign over 96.6% of the time with no apparent bias or overfitting. Using the Melatect app, users have the ability to take pictures of skin lesions (moles) and subsequently receive a mole classification. The Melatect app provides a convenient way to get free advice on lesions and track these lesions over time. A recursive computer image analysis algorithm and modified MLOps pipeline was developed to create a model that performs at a higher accuracy than existing models. Our training dataset included 18,400 images of benign and malignant lesions, including 18,000 from the International Skin Imaging Collaboration (ISIC) archive, as well as 400 images gathered from local dermatologists; these images were augmented using DeepAugment, an AutoML tool, to 54,054 images.

## I. INTRODUCTION

Malignant melanoma, referred to herein as "melanoma," is a type of cancer that in most cases starts in pigment cells (melanocytes) in the skin [1]. In this study, we omit the relevance of rare melanomas of organs such as the brain, eyes, and mouth. Melanoma is considered to be one of the deadliest skin cancers due to its rapid escalation once detected. Melanoma first shows signs in skin growths that appear discolored, strangely shaped, asymmetric, etc. [2]. There are many other characteristics of melanomas aside from common coloration and asymmetry, such as texture and existence of certain structures (clinical features), in the lesion that differentiate them from benign lesions [2]. Border definition, such as irregular, blurred, or ragged mole edges, are often signs of potential melanoma, as is the diameter of the mole; with 6 mm being the limit for a regular lesion. The most common strategy for identifying melanoma is the ABCDE strategy, where a visual inspection considering the asymmetry, borders, color, diameter, and evolution is conducted [3]. Other factors include evolution of the lesion, where the mole changes in size, shape, and color in a matter of months or years [3]. According to the American Cancer Society (ACS), about 76,380 total new cases of melanomas were diagnosed in 2016 [4]. In 2020, 100,350 new cases of melanoma were reported, indicating a 27% increase in just four years. The ACS reported 6,850 fatalities in the same year [5], a 38% decrease from 2016. These statistics suggest that early detection significantly reduces fatalities, as cases are detected more often and with greater accuracy, the corresponding fatality rate decreases in direct proportion.

Progress in terms of both computing power and storage has led to a rapid increase in assessing the potential use of artificial intelligence (AI) for various tasks. Such applications include cancer research, from going beyond the initial use by computer-aided detection (CAD) applications to including diagnosis, prognosis, response to therapy, and risk assessment. Some of the most accurate AI systems — such as face recognition features on iPhones or Google's automatic translator — have resulted from advances made in machine learning [6]. When compared to previous works, Melatect is unique because of its higher classification accuracy rates, different model development processes, continuous training pipeline, and easy-to-use iOS application.

Machine learning in healthcare has shown significant potential to transform the medical landscape ("Ascent of Machine," 2019, p. 407). Machine learning models capable of detecting breast cancer in scans have been effective as well [7]. An app that accurately classifies skin lesions would allow people to have medically assisted self exams from the comfort of their home. If the program identifies a mole as malignant, the user will be notified that the mole is potentially abnormal, and will be urged to schedule a doctor's appointment, which could help identify cancer before it advances to a stage where it can no longer be easily and feasibly resected. We recognize that such an app has to be well tested before encouraging use by the general public, which is why the app is unpublished as of yet.

## II. METHODOLOGY

### A. Overview

The modern process for getting a mole sent for a biopsy begins with a self exam where patients assess a mole and decide for themselves whether or not to consult a dermatologist. Dermoscopy, otherwise referred to as dermatoscopy or epiluminescence microscopy, is the popular method of acquiring an enlarged and detailed image of a lesion for increased clarity [8]. Dermatoscopic images of moles are generally used by dermatologists to see the nuances of the melanoma at hand. Our machine learning model aims to assist the general population, rather than dermatologists, which is why our neural network-based classification system is trained on clinical images rather than dermoscopic images [8].

### B. Melatect App Interface Overview

The Melatect app consists of a patient and clinical trials interface. The patient interface is meant for hypothetical patients to use the app as a free option for classifying melanoma. The clinical trials interface is set up so we can gather accurate data, as well as more diverse moles images from dermatologists to retrain our model. These images also retrain the machine learning model for additional accuracy once the images are verified as true malignant or benign.

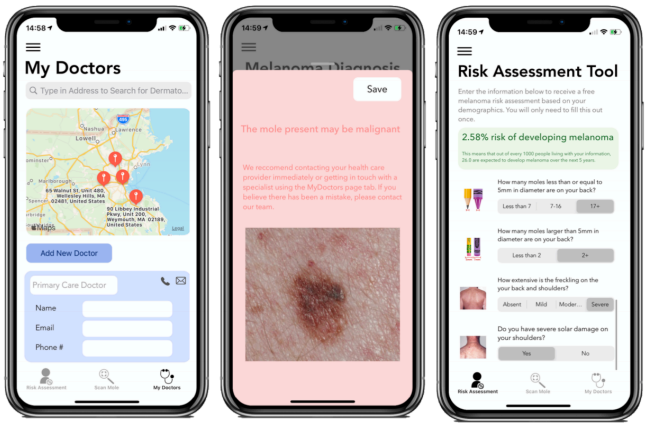

Figure 1. Patient interface screenshots with risk assessment tool, melanoma classification, and "my doctors" page.

### C. Patient Interface

The patient interface allows users to upload photos of their moles and get an immediate benign or malignant classification. The image of the mole and its corresponding parameters (classification, date, time taken, etc.) are stored as red or green markers on the 3D body model, and can be accessed at a later time. Users are notified at chosen time-intervals, requesting them to update the photos. As melanoma evolves over time, this feature tracks mole progression.

Melatect includes a risk assessment tool that outputs a percentage risk for developing melanoma based on demographics from The Melanoma Risk Assessment Tool (an open source tool developed by the National Cancer Institute for use by health professionals to estimate a patient's absolute risk of developing invasive melanoma). Melatect prompts users to take action regarding their diagnosis by searching for (using location based map feature) and contacting dermatologists from within the app.

### D. Clinical Trials Interface

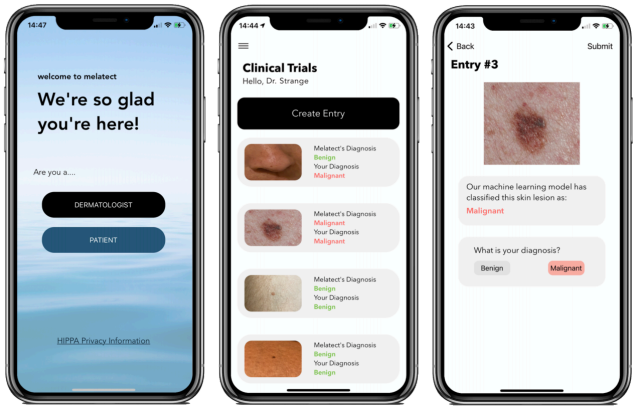

Figure 2. Clinical interface screenshots with uploading image set tool and corresponding classifications and entry page.

Our clinical trials interface (Figure 2) provides a place for dermatologists to upload photos of moles and receive a classification; Each dermatologist entry may contain up to 10 mole images and their corresponding parameters: Melatect diagnosis, dermatologist diagnosis, time and date taken. Clinical trials help us crowdsource a diverse array of mole images, used to retrain the model. This also allows us to collect additional statistics regarding the accuracy of the model.

In the future, the clinical trials interface can be modified into a diagnosis validation tool for dermatologists. For now, it is useful in updating our model to be the most accurate possible. By retraining using uploaded images, the model's total dataset continues to diversify and the accuracy continues to increase.

### III. APP DEVELOPMENT

Our project can be divided into three sections: ML Development, ML Pipeline, and App Design, each with its own plethora of development tools. For ML development, we used the Keras deep learning framework and Tensorflow backend (for the Sequential model) in Jupyter notebooks. We then conducted a model evaluation process using the Matplotlib library to graph our evaluation metrics, OpenCV computer vision library for image processing, scikit learn for model valuation, and DeepAugment for data augmentation. Meanwhile, the MLOps pipeline was set up using MLflow Model Registry and Microsoft Azure. MLflow Model Registry allowed us to store and monitor the various versions of our model and its lifecycle, and supported Azure (used as a container-based backend) as a serving endpoint.

### A. Designing iOS app

To design and develop the iOS app, we used Xcode as our IDE, and multiple Swift libraries (Azure plugins, Realm for local storage, FSCalendar/CalendarKit for mole evolution tracking, SQLite for data storage, and Lottie for animations and visual design). Location services were added for the "Contact Dermatologist" feature using the CLLocation-Manager, Core Location, and Mapkit from Xcode. The notification appearance was then customized using the Notification Content App Extension for Xcode, and the app's creative UI design was drafted and styled using Sketch and Adobe Photoshop design software.

### B. Dataset Selection

We established that the ISIC dataset would be the primary source of images for the ML model to train and test with. To increase the diversity of our dataset, we also reached out to numerous dermatologists over a period of six months, from whom we gathered a supplementary 400 images (250 benign, 150 malignant) of skin growths. This contributed to the novelty of our ML model because it increased the bounds of our dataset and used images not found in the public domain.

### IV. DATASET PREPROCESSING AND AUGMENTATION

An exploratory data analysis was performed on our dataset to better understand the nature of the data and identify patterns of predictors across classes. The data analysis was performed by determining the average image for each class, computing variability among classes (benign/malignant), and observing contrast between average images. To determine how the average benign and malignant mole looked in our dataset, the total images for each class were essentially merged into one image using Numpy along with PIL (Python Image Library), allowing us to see the gross differences that were present across the two classes. To observe which area is most variable in either class, we computed variance and standard deviation across images rather than the mean, producing two images (one from each class) that has darker or lighter areas indicating differences between malignant and benign moles.

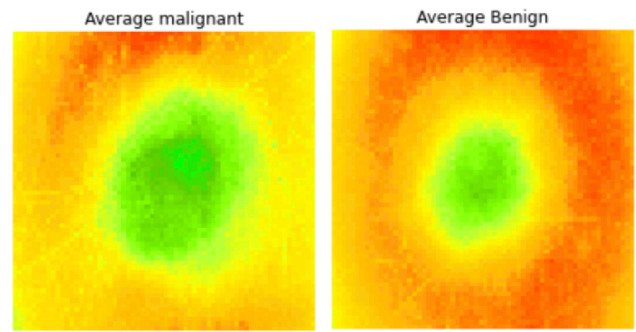

Figure 3. Average mole depictions, malignant and benign, to show differences between the classes as detected by ML model.

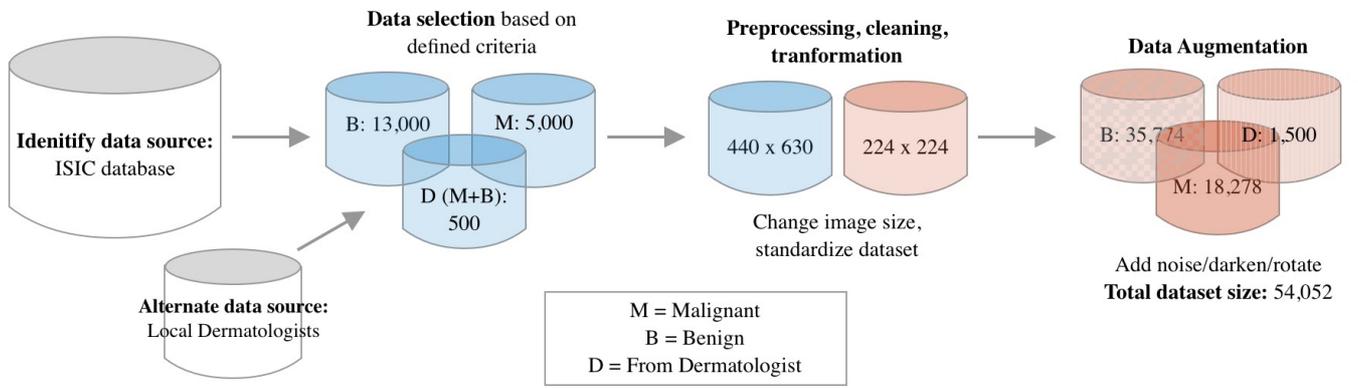

Figure 4. Data preprocessing and handling overview diagram.

Figure 3 shows the average malignant vs. benign mole heat map, depicting the differences between the average malignant mole and the average benign mole using the Matplotlib gist_rainbow colormap. From this, we were able to deduce that the comparative size, and density of growth between the two will impact the model's perception of the mole as benign or malignant. This is similar to how doctors conduct visual inspection in person - using size as an indicator. Additionally, malignant moles look more raised than benign moles, as shown in the average photos.

### A. Data Augmentation

We resized all training images to fit 224*224 to feed into the pre-trained VGG-16 model (Fig. 4). DeepAugment is an AutoML tool that reduces the error rate of CNN models [4] and is 50 times faster than Google's AutoAugment. We accounted for issues that would be encountered when taking photos of moles with our augmentation process. Augmentation strategies for each image consisted of rotating 90 degrees, additive Gaussian noise (amount: 50%, strength: 60%), and darkening the images (amount: 30%). We included an example of what two images of benign moles looked like before and after augmenting (Fig. 5).

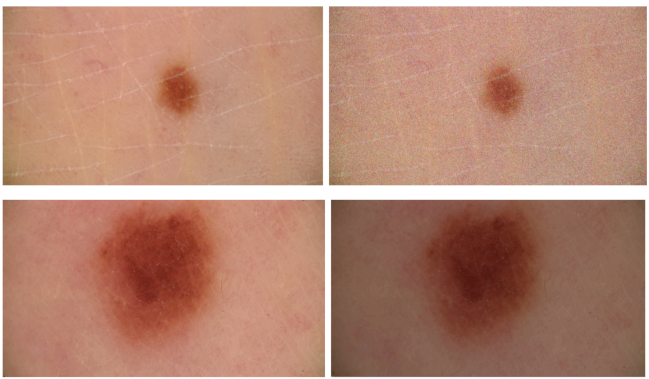

Figure 5. Noise added on the top left image to produce the right, darkening to the bottom left image to produce the bottom right.

## V. MODEL DEVELOPMENT

CNNs (convolutional neural network) have the highest accuracy for image classification problems because they can automatically detect notable features, allowing ML models to develop an advanced grasp on image data. CNN-based methods for automated machine learning have been successful for medical purposes in the past- specifically regarding tumor detection. For instance, tests with CNN have achieved results with 87% accuracy at detecting malignant tumors in breast cancer scans. There has also been research into the most efficient neural network strategies for identifying melanoma in skin growths. Dildar, et. al. [13] performed a comprehensive study on approaches for detecting melanoma using CNN, KNN (Kohonen self-organizing neural networks), ANN (artificial neural networks) and GANs (generative adversarial neural networks).

When compared, the accuracies of ANN and CNN-based approaches to identifying malignant melanoma were 88.2% and 87.5%, respectively (average of each set of trials for ANN and CNN). However, tests with ANN were done using small datasets of dermoscopic images while tests with CNN were done on the ISIC dataset using more than 400 images each. On this basis, we deduced that a CNN-based approach would be more effective for our dataset.

### A. VGG-16 Convolutional Neural Network

We used a modified VGG-16 CNN. VGG-16 has 16 total layers and was selected because it has a 92.7% top-5 test accuracy in ImageNet (dataset for 14 million images and 1000 classes); VGG-16 also performed with a top-5 classification error of 7.32% in the ILSVRC classification task in 2014, and won the localization task with a 25.32% error [9]. Compared to other CNNs tested (ResNeXt-50, Inception-v4, and AlexNet), VGG-16 produced the least overfitting (when the model fits too specifically against training data resulting in accuracy loss and inability to generalize well to new data).

We altered the original VGG-16 CNN architecture by replacing the generic softmax function (final output layer) with the sigmoid function (Fig. 6) The sigmoid function was more relevant to our training because it is intended for two class logistic regression, whereas the softmax function is intended for multiclass logistic regression [10]. Although the two produce the same output for less than two classes, using the sigmoid function results in a more lightweight ML model.

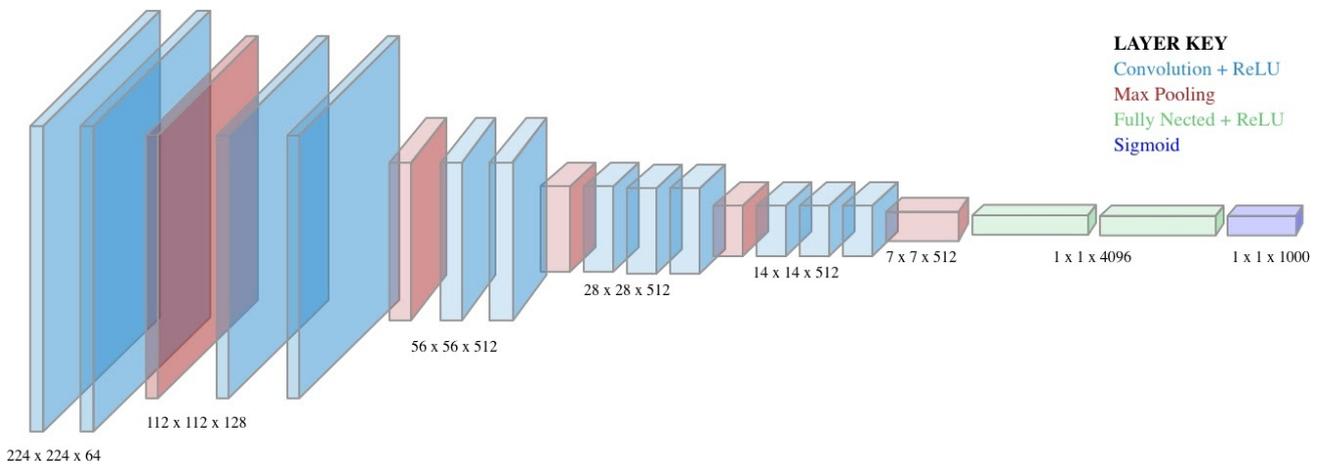

Figure 6. Depiction of VGG-16 CNN with relative position of the sigmoid function in the CNN.

### B. Transfer Learning

The VGG-16 model initiates the idea of transfer learning by utilizing models' weights for later tasks. Transfer learning is a machine learning method that uses knowledge gained from solving one problem on another, different (but related) problem. Transfer learning enables productive development by reusing a model created for a task as the starting point for another task, and has various advantages that increase the efficiency of our machine learning model including saving training time and improving baseline performance. For a task like melanoma detection, the amount of data we have post-augmentation is still considered small because of how nuanced the differences between benign and malignant moles are. With transfer learning, a ML model requires less training data because it is pre-trained. We used transfer learning to mitigate the risk of having insufficient training data and alleviate issues regarding decreased accuracy [12].

### C. Model Specifics

Our machine learning model was fine tuned using pre-trained weights from VGG-16. This was done by training the model partially, keeping the weights of the initial layers in the pre-trained VGG-16 CNN frozen while retraining only the higher layers [17]. There are 16 total layers in VGG-16, hence the name. Layers 0-14 were frozen while all subsequent layers after layer 14 were retrained. Layers up to 14 capture universal features that are easily identifiable, which were intentionally kept intact because of their relevance to the problem. While those weights are kept intact, the network is able to concentrate on learning dataset-specific features in the subsequent layers

We used stochastic gradient descent (SGD) to perform updates to the model after evaluating each batch of training samples. SGD is an optimization algorithm that estimates the error gradient for the current ML model state. The error gradient measures the direction and magnitude computed during neural network training. This value is then used to update the network weights with correct direction and magnitude.

### D. Backpropagation

In fitting a neural network for supervised learning, backpropagation is a methodology that uses gradient descent with respect to the neural network's weights to compute the gradient of the loss function (difference of current output and expected output location). Specifically, we used the binary cross-entropy loss function and 100 epochs with Adam[40] optimizer. Because our problem was binary (malignant or benign), this type of loss function was ideal.

The CNN weights were updated using backpropagation, which produces changes to each weight in order to decrease the model loss. To estimate how successfully models are performing, the cost function within backpropagation was used (defined as $c = \frac{1}{2}$ (predicted - actual) $^2$) [15] on data points fed into the network[16]. The process then begins from the lowermost layer, computing gradients with respect to the nearest weights and moving back until the first layer is reached. The correction is propagated from the cost function to the input layer (source to origin) and the effect is proportional to the responsibility of all weights [16].

### E. Dissolving Vanishing Gradient

However, there is one integral piece of our methodology that was altered based on the type of activation function we used (saturated, sigmoid). Since we used a sigmoid function at the end of the network, we standardized our dataset to eliminate the chance of forcing the last values of the gradients to 0. The chain rule, which backpropagation utilizes heavily, is based on multiplications. When the gradients dip below 1, multiplication causes the last value to be close to 0, creating a vanishing gradient [15]. A vanishing gradient can stop the training process of sigmoid or hyperbolic tangent functions (tanh). Large values create large outputs that negatively impact saturating functions like sigmoid. Therefore, images were normalized in order to force each image pixel value to be in range 0-1 or -1-1 and minimize the chance of creating a vanishing gradient [15].

## VI. MODEL EVALUATION

We collected relevant data regarding accuracy for our model on its first iteration of training (Table I).

| Before Augmentation | | | |
|---|---|---|---|
| | Benign | Malignant | Total |
| **Training (85%)** | 11262 | 4376 | 15638 |
| **Testing (15%)** | 1988 | 774 | 2762 |
| **Total** | 13000 (ISIC) + 250 (dermatologists) = 13250 | 5000 (ISIC) + 150 (dermatologists) = 5150 | 18400 |
| *After Augmentation* | | | |
| | Benign | Malignant | Total |
| **Training** | 33786 | 17504 | 51290 |
| **Testing** | 1988 | 774 | 2762 |
| **Total** | 35774 | 18278 | 54052 |

Table I. Collected data for Melatect on first iteration of training.

One of our challenges with model evaluation was an uneven distribution between the number of benign and malignant images, which would result in falsely high accuracy rates. Therefore, we specifically chose not to include classification accuracy (ratio of correct predictions to number of input samples), as it only reflects accurately for a dataset with equal number of samples for each class, and if not, results in misclassification.

### A. F1 Score

F1 score is generated by taking the mean of the precision ($\frac{true\ pos}{true\ pos\ +\ false\ pos}$) and recall $\frac{true\ pos}{true\ pos\ +\ false\ neg}$ values. The F1 score is the contribution of both, meaning that a higher F1 score is indicative of a more accurate model [22]. The equation for our F1 score is as follows, where p = precision and r = recall:

$$\frac{2 * p * r}{p + r} = \frac{2 * 0.93 * 0.947}{0.93 + 0.947}$$

If the product in the numerator dips down too low, the final F1 score decreases dramatically. A model with a good F1 score has the most drastic ratio of true:false positives as well as the most drastic true:false negatives ratio. For example, if the number of true positives to the number of false positives is 100:1, that will produce a "good" F1 score. Meanwhile, having a close ratio, say 50:51 true to false positives, will produce a low F1 score. Our F1 score indicates a fairly accurate model due to its ~94:6 true to false positives ratio.

### B. Confusion Matrix and ROC- AUC curve

Our model has an AUC of 95.23 (Fig. 7), indicating it is a generally accurate model. AUC-ROC curves are useful for understanding how accurately the model can distinguish between benign and malignant moles.

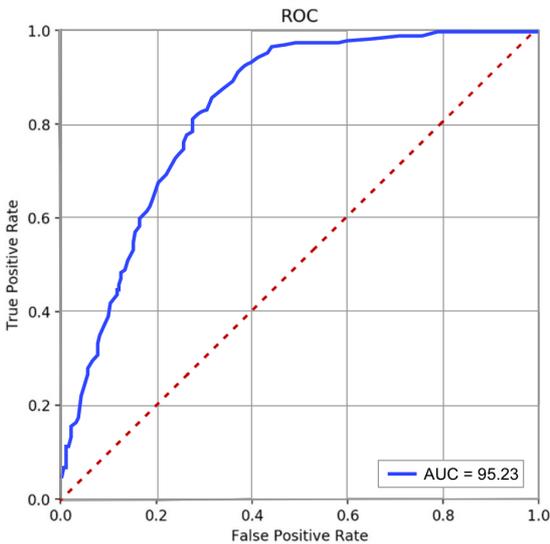

Fig 7. ROC-AUC curve displaying Area Under Curve on bottom left.

ROC is a probability curve that plots precision and recall at various classification thresholds, whereas AUC is a measurement of the whole area beneath the ROC curve, and provides a measurement for how successful the model is at discerning between classes [18]. Thus, a higher AUC correlates to a higher model performance. A model with almost a 100% accuracy rate would have an AUC of 1 (or a flawless measure of separability) and a model with no class separation ability would have an AUC of 0.5. Precision and recall are inversely proportional relationships. Therefore, decreasing the threshold would result in a higher precision and lower recall, and vice versa. From the confusion matrix, we can calculate overall model accuracy as 96.6% using $\frac{TP\ +\ TN}{TP\ +\ TN\ +\ FP\ +\ FN}$ (Fig. 8, confusion matrix).

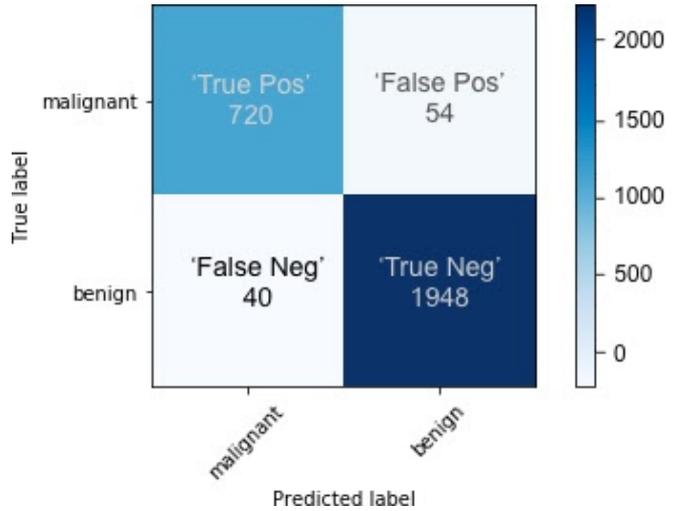

Fig 8. Confusion matrix displaying relative accuracy of different scenarios.

## VII. MODEL VALIDATION

The validation parts of the model are responsible for catching problematic behaviours or patterns (such as bias) in the input data. A model can be described as robust if its predictions stay consistently accurate, despite the features being altered due to random or unforeseen reasons. Ideally, our model should achieve the same performance every time. Robustness metrics are used to evaluate the model and determine if it should be placed into production use. If a model fails to pass the robustness benchmarks, it would be retrained with adversarial training, differential privacy, etc. so that it can pass in the next round.

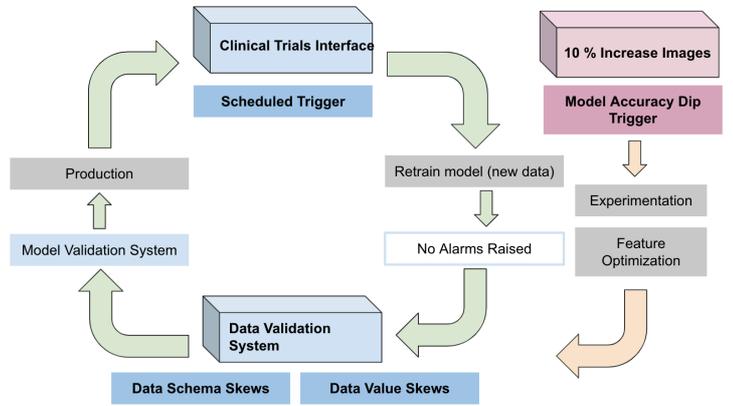

Fig 9. MLOps Pipeline and Architecture bird's eye view diagram.

### A. Flask RESTFul

Once we trained the ML model, we deployed it to a host as an API using Flask. Flask is a Python web framework that lets developers easily deploy models and generate Restful API's. When users input an image for classification, their edge device (iPhone, Android, etc.) makes a request to the host and receives a prediction. The model is automatically updated and deployed again through the pipeline every time it retrains on new data.

*B. Analyzing Bias*

To identify bias in our model, we considered features that either have a small role in the model prediction, or when they have an abnormal role. As an example, we checked our model for bias against dark skinned individuals, who have fewer training photos and play a small role in the model prediction. We used 46 images of darker skinned moles to train the model and augmented this amount to 750 images using blur, exposure, crop, and rotation techniques. Still, this only accounts for about 1.4% of our total dataset and provides for a disproportionate amount of dark to light skinned images. Due to the disproportionate amount of dark skin mole pictures given to the model to train with, we predicted there may be bias towards identifying dark skinned images as malignant. However, dark skinned individuals rarely develop melanoma and instead experience carcinoma, a type of skin cancer not seen in topical integumentary organs such as the skin [21].

We found that Melatect is able to diagnose moles as benign or malignant with equal classification accuracies and has no bias against a certain sex, age, or anatomical site of the mole. Melatect's average diagnostic accuracy is 96.6%, higher than a diagnostic accuracy for a doctor's visual inspection (77%) [20]. Because Melatect has a high classification rate, high diagnostic accuracy, and no detected biases, we believe to the extent of our knowledge that it may be useful to the public. However, we hope to conduct clinical trials to verify our accuracy rate for in-person tests.

VIII. MLOps ARCHITECTURE SUMMARY

We set up our MLOps pipeline to ensure continuous training on new, relevant training data (to improve model accuracy) and subsequently continuous deployment. The pipeline automates and manages integration, testing, releasing, and deployment in production. Automated data and model validation steps were set in place to retrain the existing model with new, collected data. Pipeline triggers were set up to retran or deploy the model systematically, rather than at random intervals. We deployed the entire pipeline, rather than simply just the trained model.

Code components in our ML pipeline are reproducible so they could be hypothetically shared across models. For example, although the Exploratory Data Analysis code would remain in its original notebook, and the source code itself was modularized.

We set up our MLOps pipeline using MLflow Model Registry and Azure. MLFlow Model Registry allowed us to store and monitor the various versions of our model and its lifecycle, and supported Azure (used as a container-based backend) as a serving endpoint.

*A. Pipeline Triggers*

We have multiple pipeline triggers that automatically retrain the model (Fig. 9) with new images aggregated from ISIC additions and clinical trials. To ensure that this new training data is usable, we set up a data validation system to check for data schema skews and data value skews. Data schema skews checks for potential irregularities in the input data that may not agree with the expected scheme (such as unanticipated or extra features) further down the pipeline steps. Data value skews refers to large statistical, pattern-like changes in the data. The pipeline will only proceed with continuous training given that there are no alarms.

After the model is trained with the new data, we have a model validation system before it is sent to production that checks for various evaluation metrics (see statistics section) against a test dataset. The new model should, in theory, perform better than the old one, and it's predictive accuracy should also be similar for all types of data (ie. it should not have a high classification accuracy for benign moles at the cost of having a low classification accuracy for malignant moles).

Our central pipeline trigger is on a schedule, retraining the model every month, given that there is at least 10% increase in images available. We also test for model performance degradation through the clinical trials interface. Given that the dermatologist logs whether the model's diagnosis was correct, we regularly can gather data about the model's accuracy in a clinical/real world setting. If the model accuracy dips below a certain point, we manually conduct some experimentation to optimize our features, and retrain the modified model.

When compared with other solutions/methodologies we researched, our skin cancer detection system is unique because of continuous training. Rather than remain at one accuracy in production and not adapt, we actively monitor the quality of the model and retrain accordingly and systemically.

IX. FUTURE WORK

We will continue to gather images for retraining the model through conducting clinical trials. Within our clinical trials, we will incorporate a metadata collection tool to understand the diversity of our data set, and improve testing for bias. We can also collect data about the app's accuracy as it pertains to real-life circumstances, through feedback forms for participating dermatologists. We will also explore the possibility of partnering with more dermatologists for official, regulated use of the app in clinics. Some features we plan to implement are a paid dermatologist consultation, diagnosis validation tool for dermatologists, and a website separate from clinical trials to aggregate images from clinics. Real-time detection will assist the user in taking precise photos of their lesion, which we will implement by creating a "skin detection model" to detect the presence of a mole or skin prior to providing a classification. We predict this will be done using some version of YOLO (You Only Look Once), an object detection algorithm meant for real-time detection. As for simplifying our MLOps pipeline, we plan to add a model-monitoring feature so users can submit feedback regarding their specific diagnoses.

X. DISCLAIMERS

Our data collection process was HIPAA compliant, as no identifying data about patients was collected in addition to the plain mole photos. The photo collection process, as well as using images from the ISIC dataset, was entirely anonymous. Multiple dermatologists were contacted regarding future steps with the app, features to be implemented, and the ethics of producing this machine learning model. As an unreleased prototype, Melatect is not available for the general public to use for diagnosis purposes or in the place of a licensed dermatologist until we acquire proper approval.

XI. ACKNOWLEDGEMENTS

We would first like to thank our advisor, Dr. Glenn Allen for providing his utmost support and advice throughout the length of this project, as well as access to computers from the MIT Kavli Institute with which we were able to run our models on, and helping us connect with dermatologists. Dr. Allen's expertise was invaluable in formulating our research paper and methodology. We are grateful to dermatologists Dr. Jessica Howie, Dr. Neal Kumar, and Dr. John Kirwood for their input during this project. We also thank the Massachusetts General Hospital Cancer Center for its assistance in providing us with clinical photos and advice on features that would benefit patients using Melatect.